\documentclass[12pt]{article}

\usepackage{amssymb,amsthm,amsfonts,psfig,epsfig,axodraw}

\def\gappeq{\mathrel{\rlap {\raise.5ex\hbox{$>$}} {\lower.5ex\hbox{$\sim$}}}}
\def\lappeq{\mathrel{\rlap{\raise.5ex\hbox{$<$}} {\lower.5ex\hbox{$\sim$}}}}

\def\beq{\begin{equation}} \def\eeq{\end{equation}} 
\def\bea{\begin{eqnarray}} \def\eea{\end{eqnarray}}

\def\bq{\begin{quote}} \def\eq{\end{quote}}

\def\nn{\nonumber}

\def\ti{\tilde}

\def\i3{\mathbb{I}}
\def\d{\delta}

\def\ra{\rightarrow}
\def\tgb{\tan \beta}

\begin{document} 

\pagestyle{empty} 
\begin{flushright}   ROMA1-TH/1358-03 \\ SACLAY-T03/114  \end{flushright}  
\vskip 2cm    
\def\thefootnote{\fnsymbol{footnote}}

\begin{center}  
{\large \bf Heavy Triplets: Electric Dipole Moments vs Proton Decay } \vspace*{5mm}    
\end{center}  
\vspace*{5mm} \noindent  \vskip 0.5cm  

\centerline{\bf Isabella Masina$^{(a)}$ and Carlos A. Savoy$^{(b)}$} 
\vskip 0.5cm  
\centerline{\em (a) Centro Studi e Ricerche "E. Fermi", Via Panisperna 89/A}
\centerline{\em I-00184 Roma, Italia }
\vskip 0.3 cm
\centerline{\em (b) Service de Physique Th\'eorique 
\protect\footnote{Laboratoire de la Direction des Sciences de la Mati\`ere du Commissariat \`a l'\'Energie
Atomique et Unit\'e de Recherche Associ\'ee au CNRS (URA 2306).} , CEA-Saclay}
\centerline{\em F-91191 Gif-sur-Yvette, France} \vskip2cm   

\centerline{\bf Abstract}  
The experimental limit on the electron electric dipole moment 
constrains the pattern of supersymmetric grand-unified theories with right-handed neutrinos. 
We show that such constraints are already competing with the well known ones
derived by the limit on proton lifetime. \\
\vskip 0.3cm 
%\noindent PACS: 12.10.Dm, 12.60.Jv, 11.10.Hi, 13.40.Em.

\vspace*{1cm} \vskip .3cm  \vfill   \eject  
\newpage  

\setcounter{page}{1} 
\pagestyle{plain}

\def\thefootnote{\arabic{footnote}}
\setcounter{footnote}{0}

%%%%%%%%%%%%%%     Introduction      %%%%%%%%%%%%%%%%%%%%%%%%%%%%%%%%%%

%\section{Introduction}

The experimental limit on the proton lifetime $\tau_p$ \cite{exp-p} represents a
crucial test \cite{pdec-lit} for supersymmetric grand unified theories (GUTs). 
In particular, the minimal $SU(5)$ version is ruled out \cite{gotonihei, muray} 
-- unless particular sfermion mixings are assumed \cite{senj} -- 
because the experimental limit on the decay mode $p \ra K^+ \bar \nu$ implies 
a lower limit on the triplet mass which, for sparticle masses up to a few TeV, 
is much higher than the value demanded for gauge coupling unification \cite{couplunif}.
Supersymmetric GUT models where $\tau_p$ remains consistent with experiment 
usually exploit the presence of two or more massive colour triplets with a peculiar mass matrix 
structure \cite{PD-lit}.

The experimental limit on the electron electric dipole moment $d_e$ \cite{exp-de} 
also provides interesting constraints on supersymmetric GUTs with heavy right-handed neutrinos.  
Indeed, the radiative effects from the colour triplets and neutrino Yukawa couplings 
could give rise to sizeable contributions to $d_e$, recently calculated in \cite{isa-edm} 
\footnote{For the pure seesaw case, see {\it e.g.} \cite{romstr, ellis-edm, isa-edm}.}.  
Besides their dependence on supersymmetric masses, 
these contributions are basically proportional to 
$\log( \Lambda / M_T)$ and $\log( \Lambda / M_R)$, where $M_T$ and $M_R$ 
stand for the triplet and the right-handed neutrino masses respectively, and to a 
combination of neutrino and triplet Yukawa couplings. 
Then, once the triplet Yukawa couplings and the seesaw parameters are assigned, 
the experimental upper bound on $d_e$ translates into an upper bound on
$\log(\Lambda / M_T)$, whose dependence on sparticle masses will be shown in the following.  

Due to the many parameters involved, $\tau_p$ and $d_e$ represent
{\it complementary} tests for supersymmetric GUTs endowed with the seesaw mechanism. 
Notice that, before the experimental limit on $\tau_p$ \cite{exp-p} could be significantly improved, 
planned experiments are expected to strengthen the present limit on $d_e$ 
by three \cite{expf-de} to five \cite{lam} orders of magnitude.
Within this context, the aim of this letter is:

\noindent A) To show that {\it  in supersymmetric GUT models with
right-handed neutrinos, the present constraints from $d_e$ \cite{exp-de} are already
competitive with $\tau_p$ ones \cite{exp-p}.} 
This can be done, for definiteness, in the context of the minimal $SU(5)$ model
by comparing the $d_e$ experimental limit 
with the $d_e$ upper prediction calculated by using the lower limit on $M_T$ from $\tau_p$ searches.
Indeed, we find that such a prediction exceeds the $d_e$ experimental limit 
even for quite small neutrino Yukawa couplings and moderate values of $\tgb$.
This means that also in more realistic GUT models one should always check the consistency 
with the experimental limit on $d_e$ -- and not only with that on $\tau_p$;

\noindent B) To show that {\it supersymmetric GUT models consistent with the $\tau_p$ experimental limit 
can violate the limit on $d_e$.} 
Potentially realistic GUT models generically have two or more massive triplets.
While the proton decay rate could be reduced down below the experimental limit
as a consequence of the triplet mass matrix structure,
$d_e$ is quite insensitive to the latter and, rather, it basically increases with the number of states involved 
in the radiative corrections.  
As a case study, we consider an $SO(10)$ model with one \underline{$10$} to give up-quark
and neutrino masses and another \underline{$10$} to give down-quark and
charged lepton masses, and moderate values of $\tgb$. 
When triplets are roughly degenerate at $M_T=O(10^{17})$ GeV, both  $\tau_p$ and $d_e$ strongly violate the
experimental limit.
Instead, with a pseudo-Dirac structure for the triplet masses, the $\tau_p$ bound is easily evaded 
while $d_e$ is marginally affected and remains in conflict with experiment;

\noindent C) To discuss {\it the size of neutrino Yukawa couplings 
such that the limits on $M_T$ from $d_e$ and from $\tau_p$ are of comparable magnitude}.
It turns out that for moderate $\tgb$ this already happens with rather small Yukawa couplings if the relevant
sparticles lie below the TeV region.  
We display a comparison with several classes of seesaw models and we provide
some comments on related processes such as $\mu \ra e \gamma$.

Planned searches for $d_e$ would have a strong impact on the conclusions of the present
analysis, which would be considerably strenghtened. 
Thus, it is worth both to stress the r\^{o}le of $d_e$ as a test for supersymmetric GUT models
and to calculate it in the context of explicit examples.

%%%%%%%%%%%%%%% The minimal %%%%%%%%%%%%%%%%%%%%%%%%%%%

\noindent \underline{ \large A) $d_e$ {\it vs} $\tau_p$ \textit{with one massive triplet}}

Let us first consider the case of one triplet-antitriplet pair, $H_{3u}$ and $H_{3d}$, 
which can be accomodated, together with  the two electroweak symmetry breaking Higgs doublets, 
into $H_5 = (H_{3u}, H_{2u})$ and $\bar H_{5} = (\bar H_{3d}, \bar H_{2d})$, transforming 
in a \underline{$5$} and a \underline{$\bar 5$} of $SU(5)$, respectively.
Their Yukawa couplings to matter and their masses in the superpotential
are denoted as follows:
\bea
{\cal W} & \ni &  
Q^T A Q H_{3u} + {U^c}^T  B E^c H_{3u} +Q^T  C L \bar H_{3d} + {U^c}^T D D^c \bar H_{3d} 
+ {N^c}^T  E D^c H_{3u} \nn\\
&{}& + {U^c}^T y_u Q H_{2u} + {D^c}^T y_d Q \bar H_{2d} + {E^c}^T y_e L \bar H_{2d} + {N^c}^T y_\nu L H_{2u}  \nn\\
&{}& + \frac{1}{2} {N^c}^T M_R N^c + \bar H_{3d} M_T H_{3u} + \bar H_{2d} \mu  H_{2u}~~ .
\label{wtriplets}
\eea
\noindent The minimal $SU(5)$ relations are:
\beq
y_u  = y_u^T= - 2  A =  B~~~~~~~~~~~~~  y_e = y_d^T = - C =  D    .
\label{minsu5}
\eeq 
while in the minimal $SO(10)$ with two \underline{$10$}'s the additional relation $y_u = y_\nu$ holds. 
In non minimal scenarios, these relations are affected by non renormalizable operators 
in the superpotential. 
All the $B$, $L$ and $CP$ violating effects considered in this paper originate
from the parameters in the superpotential (\ref{wtriplets}).

We indicate with $\widehat{ }$ a real and diagonal matrix and
we conveniently work (at all scales) in the basis where $y_e= \hat y_e$, $y_d = \hat y_d$ and $M_R=\hat M_R$ 
so that the unitary matrices which diagonalise $y_u$ and $y_\nu$ 
encompass all the flavour and CP violating parameters:
$y_u= \phi V_{CKM}^T \hat y_u \psi_u V_{CKM} \phi$ where $V_{CKM}$ 
is the CKM matrix in the standard parametrization, $\psi_u \equiv {\rm diag}(e^{i \psi_1}, e^{i \psi_2}, 1)$ and 
$\phi \equiv {\rm diag}(e^{i \phi_1}, e^{i \phi_2}, 1)$; 
$y_\nu = V_R \hat y_\nu V_L$, where $V_L$ has a CKM-like parameterization 
while $V_R$ is a general unitary matrix (with 6 phases). 

The supersymmetric contributions to $d_e$ depends on slepton masses, mixings and phases. 
We adopt here the following conventions for the $3 \times 3$ slepton mass matrices (up to lepton mass terms), 
consistently defined in the lepton flavour basis where the charged lepton mass matrix, $m_{\ell}$, is diagonal:
\beq
 {\tilde \ell}_L^\dagger m_L^2( \mathbb{I}+ \delta ^{LL} ) {\tilde \ell}_L +
{\tilde \ell}_R^\dagger m_R^2 ( \mathbb{I} + \delta ^{RR} ) {\tilde \ell}_R  \\ 
+ [ {\tilde \ell}_L^\dagger (  (a_e^*  - \mu \tan\beta) \hat m_{\ell}  + m_L m_R \delta ^{LR}  ) {\tilde \ell}_R 
+ h.c.] 
\label{sleptonm2}
\eeq
\noindent where $a_e$ is the average lepton $A-$term, and $m_L$ and $m_R$ are average masses 
for L and R sleptons, respectively. 
Since the present experimental bounds on lepton flavour violating (LFV) decays and EDMs
already point towards family blind soft terms with very small diagonal CP violating phases, 
at the scale $\Lambda= M_{Pl}$ 
we assume the mSUGRA boundary conditions, namely all  $\d$ matrix
elements vanish and  
$m^2_L = m^2_R = m_0^2 $, $a_e = a_0$, with real $m_0$, $a_0$, $\mu$-term.
We also assume universal real masses $\ti M_{1/2}$ for the gauginos, 
consistently with grand unification. 
In this paper we assume that at lower scales these $\d$'s are generated by the RGE 
evolution of the soft parameters.
The results obtained in the mSUGRA framework generalize
(at least to a large extent) to other models.

Under these circumstances, it has been pointed out \cite{isa-edm}
that when a couple of triplets and right-handed neutrinos are simultaneously present, 
the most important amplitude for $d_e$ is the one involving
double insertions of flavour non-diagonal $\delta$'s 
(although \cite{isa-edm} focussed on minimal $SU(5)$, we check here that this is a general result).  
The contributions from the heavy triplet and the right-handed neutrino states to the non-diagonal
entries of the $\d$'s are at lowest order (always in the basis where $y_e$ is diagonal):
\bea
\d^{RR} & = & - \frac{1}{(4 \pi)^2}\frac{3 m_0^2 + a_0^2}{m_R^2} 
(6 ~B^T \ell_T B^*  ) \nn\\
\d^{LL}  & = &  - \frac{1}{(4 \pi)^2} \frac{3 m_0^2 + a_0^2}{m_L^2} 
(6 ~C^\dagger \ell_T  C + 2 ~y_\nu^\dagger \ell_{\hat M} y_\nu ) \\
\d^{LR} & = & - \frac{1}{(4 \pi)^2}  \frac{a_0}{m_L m_R}
(6~\hat m_\ell B^T \ell_T B^*  + 6~C^\dagger \ell_T C \hat m_\ell 
+ 2~y_\nu^\dagger \ell_{\hat M} y_\nu \hat m_\ell) \nn
\eea
where all the Yukawa couplings are defined at $\Lambda$ and
\beq
\ell_T \equiv \ln(\Lambda/M_T) \quad \ell_{\hat M_i} \equiv \ln(\Lambda/M_i)~~,
\label{ln}
\eeq
the diagonal matrix $\ell_{\hat M}$ accounting for a possible hierarchy in the right-handed neutrino spectrum. 
The non-diagonal $|\d_{ij}|$'s induce -- and are constrained by 
(see \cite{sleptonarium} for a recent analysis) -- 
the LFV decays $\ell_i \ra \ell_j \gamma$. 
Then, defining ${\cal C} \equiv y_\nu^\dagger \ell_{\hat M} y_\nu$, such limits on the $|\d_{ij}|$'s 
also provide limits on $|{\cal C}_{ij}|$'s \cite{imsusy02}.

Omitting terms that are less relevant or higher order in the $\d$'s matrix elements,
and working in the mass insertion approximation as in ref. \cite{sleptonarium}, 
the most important contribution to $d_e$ reads:
\beq
d_e  =  \frac{3 e \alpha \tilde M_1}{(4 \pi)^5 |\mu |^2 \cos^2 \theta _W} 
~{\cal{I}}_{11} ~\left(
(\mu \tgb -a_0) \frac{ \bar M_0^4 }{m_R^2 m_L^2} ~ I''_B 
+ a_0 ( \frac{\bar M_0^2} {m_L^2}~I'_{B,L}+ \frac{\bar M_0^2}{m_R^2}~I'_{B,R})   
\right)~~, 
\label{eqde}
\eeq 
\noindent with
\beq 
{\cal I}_{11}  = {\cal I}m \left( ( {\cal C} + 3 C^\dagger \ell_T C ) ~\hat m_\ell~ B^T \ell_T B^*) \right)_{11}
\label{Irad} 
\eeq
\noindent where $\tilde M_1$ is the bino mass, $\bar M_0^2 \equiv 3 m_0^2+a_0^2$ 
and the functions $I''_B,\ I'_{B,R},\ I'_{B,L}$ of the sparticle masses are defined 
in \cite{sleptonarium} where approximations are also provided. For instance, 
when $m_R^2 \approx m_L^2 \equiv \bar m^2$   
this gives the order of magnitude estimate: 
\beq
d_e \approx \left( 2 \times 10^{-26} \mathrm{e~ cm} \right) ~
\frac{\tilde M_1}{\bar m} h_1(\frac{\tilde M_1^2} { \bar m^2}) ~ \frac{{\rm TeV}^2}{\bar m^2} 
~~\frac{ \mu \tgb ~{\cal I}_{11}}{\bar m ~m_\tau}  
\eeq
\noindent with $h_1(x)$ given in \cite{sleptonarium} and such that $0.1 < \sqrt{x}h_1(x) < 0.2$ for 
the reasonable range $0.02 < x < 3$.

If $C^\dagger C$ does not deviate too much from the minimal condition (\ref{minsu5}), 
$C^\dagger C \approx \hat y_e^\dagger \hat y_e$ and the corresponding term in (\ref{Irad}) is 
negligible.  
Analogously, 
$B^T B^*$ is expected to be close to $y_u^T y_u^* = \phi V_{CKM}^T \hat y_u^2 V_{CKM}^* \phi^*$.
Then, defining $V_{td} \equiv |V_{td}| e^{i \beta}$, 
${\cal C}_{31} \equiv |{\cal C}_{31}| e^{i \phi_{{\cal C}_{31}}}$, 
the dependence from the relevant neutrino Yukawas in (\ref{Irad}) can be made explicit
\footnote{Eq. (\ref{dep}) holds up to corrections coming from the term proportional to $m_\mu$
and which naturally are of $O(10^{-3} |{\cal C}_{21}|/|{\cal C}_{31}|)$.}: 
\beq
{\cal I}_{11}  \approx - m_\tau y_t^2  |V_{td}| |V_{tb}| |{\cal C}_{31}|  
\sin(\underbrace{\beta+\phi_{{\cal C}_{31}}+\phi_1}_{\large \equiv \phi_{d_e}}) ~\ell_T~~.
\label{dep}
\eeq
Notice that the combination of CP phases $\phi_{d_e}$
mixes the known phase of the quark sector $\beta$ 
with that of the neutrino sector $\phi_{{\cal C}_{31}}$ and the phase $\phi_1$ which
becomes unphysical when $SU(5)$ is broken.
Therefore, the $d_e$ dependence on the seesaw parameters is in $|{\cal C}_{31}|$.  
As already mentioned, $|{\cal C}_{ij}|$ would also induce $\ell_i \rightarrow \ell_j \gamma$ but,  
while the experimental limits do provide interesting upper bounds on
$|{\cal C}_{21}|$ and - to some extent - on $|{\cal C}_{32}|$ (see e.g. \cite{imsusy02}),
they are too weak to constrain $|{\cal C}_{31}|$ at the level corresponding to perturbative 
Yukawa couplings. 

For sufficiently hierarchical $y_{\nu}$ eigenvalues, 
$|{\cal C}_{31}| \approx y_{\nu 3}^2 |{V_L}_{\tau 1}| |{V_L}_{\tau 3} | (V_R^\dagger \ell_{\hat M} V_R )_{33}$ 
and $\phi_{{\cal C}_{31}} \approx \beta _L$, which is the equivalent of $\beta$ in $V_L$, namely
${V_L}_{\tau 1} \equiv |{V_L}_{\tau 1}| e^{i \beta_L}$.
Therefore, the eventual dependence in (\ref{dep}) on the $V_R$ phases 
- related to the phase relevant for leptogenesis - is suppressed in favour of $\beta _L$. 

In $SO(10)$ inspired models, for instance, neutrino eigenvalues are hierarchical with
$y_{\nu 3} \approx y_t$ and one also expects $|{V_L}_{\tau 1}| \approx |V_{td}|$ and $\beta _L \approx \beta$,
yielding \footnote{In this case,
to naturally reproduce light neutrino masses, $V_R$ is expected to have small mixings.}
$|{\cal C}_{31}| \approx 0.05$ and $\phi_{d_e} \approx (50^{\circ} + \phi_1)$.  
In such a framework, this can be considered as an estimate of $|{\cal C}_{31}|$ on the low side, 
but other models prefer $|{\cal C}_{31}| \sim O(1)$. 
We postpone the discussion of the detailed predictions for ${\cal C}_{31}$ and $d_e$ 
in different classes of neutrino mass models to the third part of this letter where we address the issue C).
However, as discussed later on, when there are more massive triplets - as in potentially realistic
$SO(10)$ models - the overall numerical coefficient in (\ref{Irad}) gets enhanced due to the larger number 
of states involved in the RGE.

\begin{figure}[!h]
\vskip .2 cm
{
\psfig{file=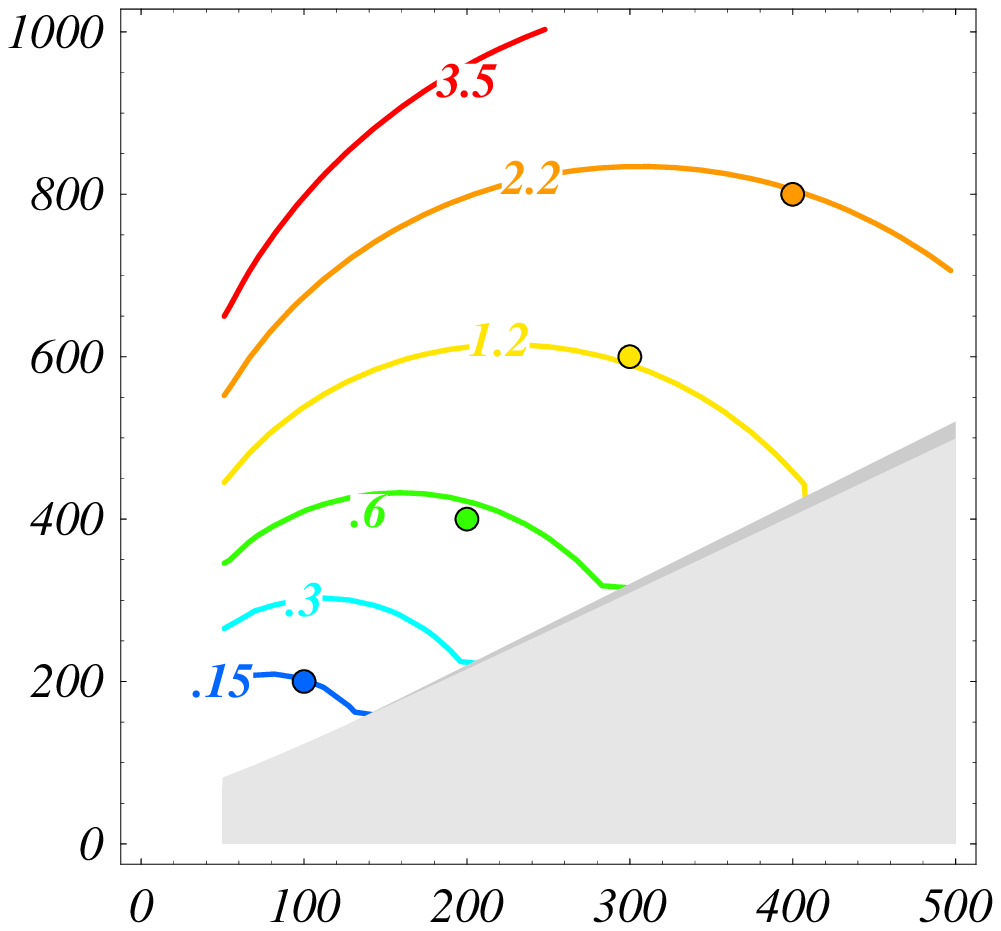,width=.5 \textwidth}
\put(-210,220){a) Upper limit on {\Large $\log_{10} (\frac{M_{Pl}}{M_T})$ } from $d_e$}
\put(-230,155){ $m_R$ }  \put(-235,140){ (GeV)}  
\put(-120,-10){ $\ti M_1$ (GeV) } 
\put(-150,65){ $P_1$ } 
\put(-115,100){ $P_2$ } 
\put(-75,130){ $P_3$ } 
\put(-40,165){ $P_4$ } 
\put(-130,40){ $ \times \frac{3}{\tgb} \frac{0.2}{|{\cal C}_{31}| \sin\phi_{d_e}    }  \frac{d_e [{\rm e ~cm}]}{10^{-27}} $}
}~~~~
{
\psfig{file=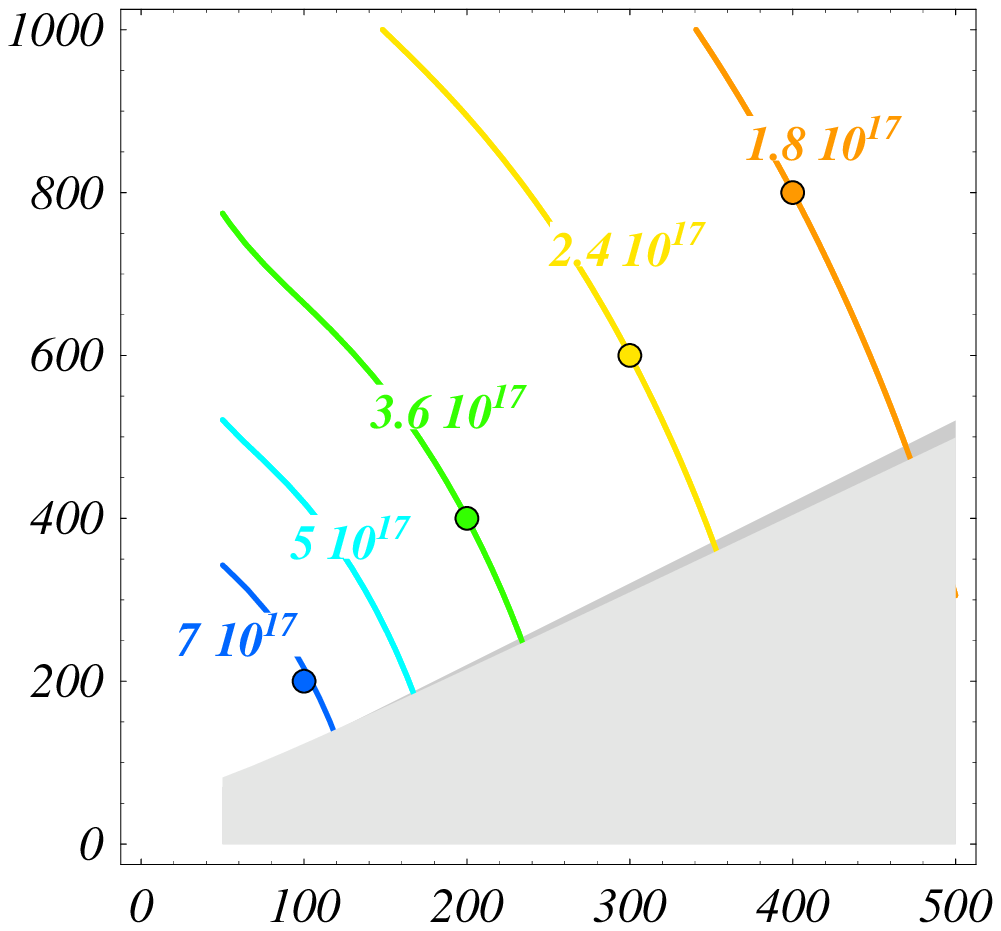,width=.5 \textwidth}
\put(-200,220){b) Lower limit on {\Large $M_T$} from $p \ra K^+ \bar \nu$}
\put(-230,155){ $m_R$ }  \put(-235,140){ (GeV)}  
\put(-120,-10){ $\ti M_1$ (GeV) } 
\put(-150,65){ $P_1$ } 
\put(-115,100){ $P_2$ } 
\put(-75,130){ $P_3$ } 
\put(-40,160){ $P_4$ } 
\put(-100,40){ $ \times  \sqrt{ \frac{\tau_p [{\rm yrs}]}{1.9~10^{33}}} $} }
\caption{a) Upper limit on $\log_{10} (M_{Pl}/M_T)$ in minimal $SU(5)$. 
It is inversely proportional to the reference values $ \tgb =3 $, 
$|{\cal C}_{31}|  \sin \phi_{d_e} =0.2$, $d_e < 10^{-27}$ e cm.
b) Lower limit on $M_T$ from the proton decay mode $p \ra K^+ \bar \nu$. 
We have taken $\tgb=3$, $A_3 =1.32$ and $A_s=0.93$  \cite{couplunif}, $-\alpha = \beta = 0.014$ GeV$^3$ 
\cite{elmadr}.}
\label{F_de-p}
\end{figure}

Once the supersymmetric masses are specified, 
one can extract an upper limit on $\log(\Lambda / M_T)$ from the experimental limit on $d_e$
which is inversely proportional to $\tgb ~|{\cal C}_{31}| \sin \phi_{d_e}$.  
This is displayed in fig. \ref{F_de-p}a) in the plane $(\ti M_1, m_R)$. 
In this plot we take the mSUGRA constraints, with $a_0^2 = \ti M_{1/2}^2 + 2 m_0^2$ 
and $\mu$ is fixed by e.w. symmetry breaking. 
We assume the relations (\ref{minsu5}) for the Yukawa couplings. 
Also shown are 'benchmark' points $P_i$ in the supersymmetric parameter space
for later use.

Let us now compare the previous limit with the limit on $M_T$ from the bounds on $\tau_p$. 
Integrating out the colour triplet one obtains the baryon number violating superpotential:
\beq
w_{eff}= Q^T A Q \frac{1}{M_T} Q^T C L + {U^c}^T B E^c 
\frac{1}{M_T} {U^c}^T D D^c ~~.
\label{weff}
\eeq
\noindent The relevant effective operators for proton decay are obtained from the two terms in (\ref{weff}) 
by the additional exchange of a wino and a higgsino, respectively. 
For large $\tgb$, the most important graph is the higgsino one \cite{gotonihei}, 
but for $\tgb \lesssim 10$ the amplitude with wino dressing cannot be neglected.
With one massive triplet, one obtains a lower limit on its mass $M_T$ 
which depends on appropriate combinations of the triplets couplings $ A,  B$, $C,  D$, 
on the sparticle masses and on the hadronic matrix elements \cite{elmadr}.  

In minimal $SU(5)$, only the supersymmetric parameters, including $\tgb$, and the two phases $\psi_1, \psi_2$ 
remain as free parameters. The higgsino amplitude alone is insensitive to $\psi_1, \psi_2$. 
However for $\tgb \lesssim 10$, the wino comes into play and the prediction for
proton lifetime varies up to one order of magnitude with $\psi_1, \psi_2$.  
In fig. \ref{F_de-p}b), the minimal $SU(5)$ limits on $M_T$ are shown for $\tgb = 3$ and values of
$\psi_1$ and $ \psi_2$ that maximize $\tau_p$. 
Notice that for lighter sparticles, the limits from $d_e$ already compete with those from $\tau_p$.

\begin{figure}[!b]
\vskip .2 cm
\centerline{
\psfig{file=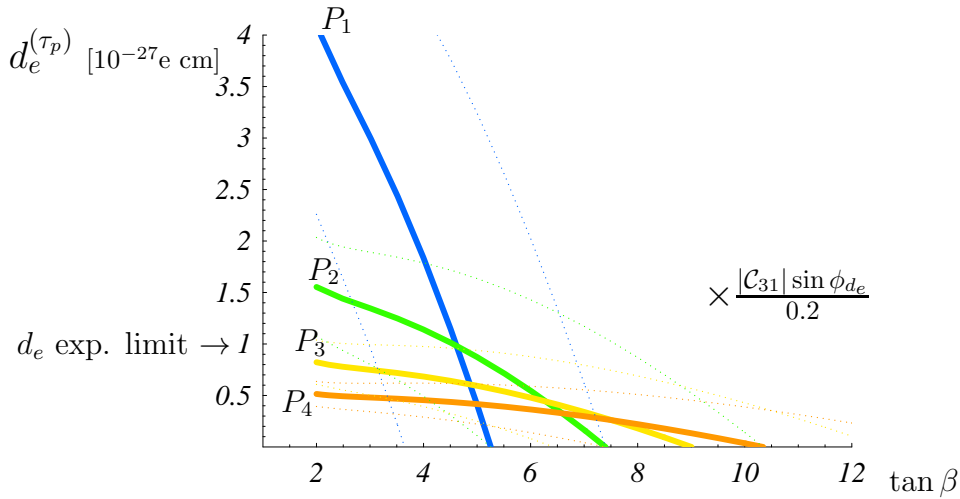,width=.6 \textwidth}
\put(-333,160){ \large $d_e^{(\tau_p)}$ \footnotesize $ [10^{-27} {\rm e~ cm}]$} 
\put(-215,175){ $P_1$ } 
\put(-220,80){ $P_2$ } 
\put(-225,53){ $P_3$ } 
\put(-230,30){ $P_4$ } 
\put(-330,52){ $d_e$ exp. limit $\ra$  } 
\put(0,0){ $\tgb$  } 
\put(-70,70){ \large $ \times  \frac{ |{\cal C}_{31}| \sin \phi_{d_e}}{0.2} $} }
\caption{$d_e^{(\tau_p)}$ is the
maximum allowed value for $d_e$, obtained by using the lower limit on $M_T$ from $\tau_p$,
for the supersymmetric parameters defined by points $P_i$ in fig. \ref{F_de-p};
it is proportional to $|{\cal C}_{31}| \sin \phi_{d_e}$, which is taken to be $0.2$ in the plot. 
We take the minimal $SU(5)$ relations for the triplets couplings (\ref{minsu5}),
$\Lambda = M_{Pl}$ and the $\psi_1$, $\psi_2$ values which maximize proton lifetime. 
Dotted lines show the effect of an uncertainty by a factor $1/2$ and $2$ on the lower limit 
on $M_T$ from $\tau_p$. }
\label{Fde_comb}
\end{figure}

A way to make this comparison more direct, is to compare the experimental limit on $d_e$
with $d_e^{(\tau_p)}$, defined as the maximum allowed value for $d_e$ calculated according to eqs. 
(\ref{eqde}), (\ref{Irad}) and plugging in the lower limit on $M_T$ provided by the experimental limit on $\tau_p$. 
This is shown in fig. \ref{Fde_comb} as a function of $\tgb$ for the various points $P_i$;  
$d_e^{(\tau_p)}$ falls down at the value of $\tgb$ where the lower limit on $M_T$ from $\tau_p$ 
approaches $M_{Pl}$.
For $|{\cal C}_{31}| \sin \phi_{d_e} \sim 0.2$, $d_e^{(\tau_p)}$ is larger than the experimental limit
in a region of moderate values of $\tgb$ and lower sparticle masses. 
Hence, in that region the experimental limit on $d_e$ is competitive with the $\tau_p$ one.
This performance of $d_e$ quickly increases with the (theoretical) input for 
$|{\cal C}_{31}| \sin \phi_{d_e}$ and, of course, with any improvement on the $d_e$ experiments.

Figs. \ref{F_de-p} and \ref{Fde_comb} were derived assuming the minimal $SU(5)$ relations (\ref{minsu5}). 
While $\tau_p$ depends on several first generations Yukawa couplings and mixings,  
from (\ref{dep}) and the following discussion that $d_e$ depends on the contrary on those of 
the heaviest generation. 
Then, by relaxing (\ref{minsu5}), $d_e$ should not change a lot while $\tau_p$ might do so.

%%%%%%%%%%%%% non min: piu' tripletti %%%%%%%%%%%%%%

\noindent \underline{ \large B) $d_e$ {\it vs} $\tau_p$ \textit{with more than one triplet}}

With one massive triplet, 
it looks quite unnatural to fulfill the 
experimental constraints on
$\tau_p$ 
through {\it ad hoc} sets of triplet Yukawa couplings. 
On the contrary, it is well known that with more triplets 
and appropriate structures for their mass matrix $M_T$ \cite{PD-lit},
$\tau_p$ can naturally exceed the experimental limit. 
Because in this case $\tau_p$ and $d_e$ put bounds on 
different combinations of triplet masses, the above direct comparison is not possible anymore
and one has rather to consider specific GUT models. 
Indeed, while $\tau_p$ results from the interference between the many amplitudes with triplet exchange, 
for $d_e$ all the heavy states contributions add up in the RGE calculation of the $\d$'s.
To establish whether, with more triplets, $d_e$ remains competitive (not just complementary) to $\tau_p$,
one must check that in models where the structure of $M_T$ allows to escape the $\tau_p$ limits,
those from $d_e$ are not simultaneously evaded. 
We show why this is actually the case by studying in some detail a typical example.

Let us consider a minimal version of $SO(10)$ with two \underline{10}'s of Higgs fields.
In the basis where Higgs doublets are diagonal, they are denoted by  indices $u$ and $d$, 
since they couple respectively to matter fields with the (symmetric) Yukawa coupling matrices 
$y_u$ and $\hat y_d$.
In this family basis, also
$\hat y_e =\hat y_d$ and $y_\nu=y_u= \phi^{1/2} V_{CKM}^T \hat y_u \psi_u V_{CKM} \phi^{1/2}$.  
Decomposing the $\underline{10}$'s
\footnote{ 
$10_{u} \equiv H_{5u}[H_{3u} , H_{2u}] + \bar H_{5d}[\bar H_{3u} , \bar H_{2u}]$, 
$10_{d} \equiv H_{5d}[H_{3d} , H_{2d}] + \bar H_{5d}[\bar H_{3d} , \bar H_{2d}]$. }
into the electroweak
$\underline{2}$'s and colour $\underline{3}$'s, their mass matrices are denoted as:
\beq
(\matrix{  \bar H_{2d} &  \bar H_{2u}} )  
\left( \matrix{ \mu &  0 \cr 0 & M_H} \right)
\left( \matrix{ H_{2u} \cr  H_{2d} } \right)  \quad \quad \quad
(\matrix{ \bar H_{3d} &  \bar H_{3u} }) ~
M_T
\left( \matrix{  H_{3u} \cr  H_{3d} } \right)
\label{mdmt}
\eeq
where $\mu$ is the $O({\rm e.w. })$ supersymmetric mass of the light
doublets ${H}_{2u}$, $\bar H_{2d}$ getting non zero v.e.v.'s. 
In the following, the eigenvalues of the matrix $M_T$ will be referred to as $M_{T_1}, M_{T_2}$.
Everything is thus known from low energy observables but $\phi$, $\psi_u$, $M_R$ and $M_T$. 
To maintain the notation used until now,
it is convenient to redefine $y_\nu$, $y_u$ in the basis where $M_R$ is diagonal so that $y_\nu$ 
becomes $y_\nu= V_{R} \hat y_u  V_{CKM}$ while $y_u= \phi V_{CKM}^T \hat y_u \psi_u V_{CKM} \phi$. 
Hence, in this model ${\cal C}_{31} \approx 0.05 e^{i \beta}$. 

\begin{figure}[!h]
\vskip 1cm
\centerline{
\psfig{file=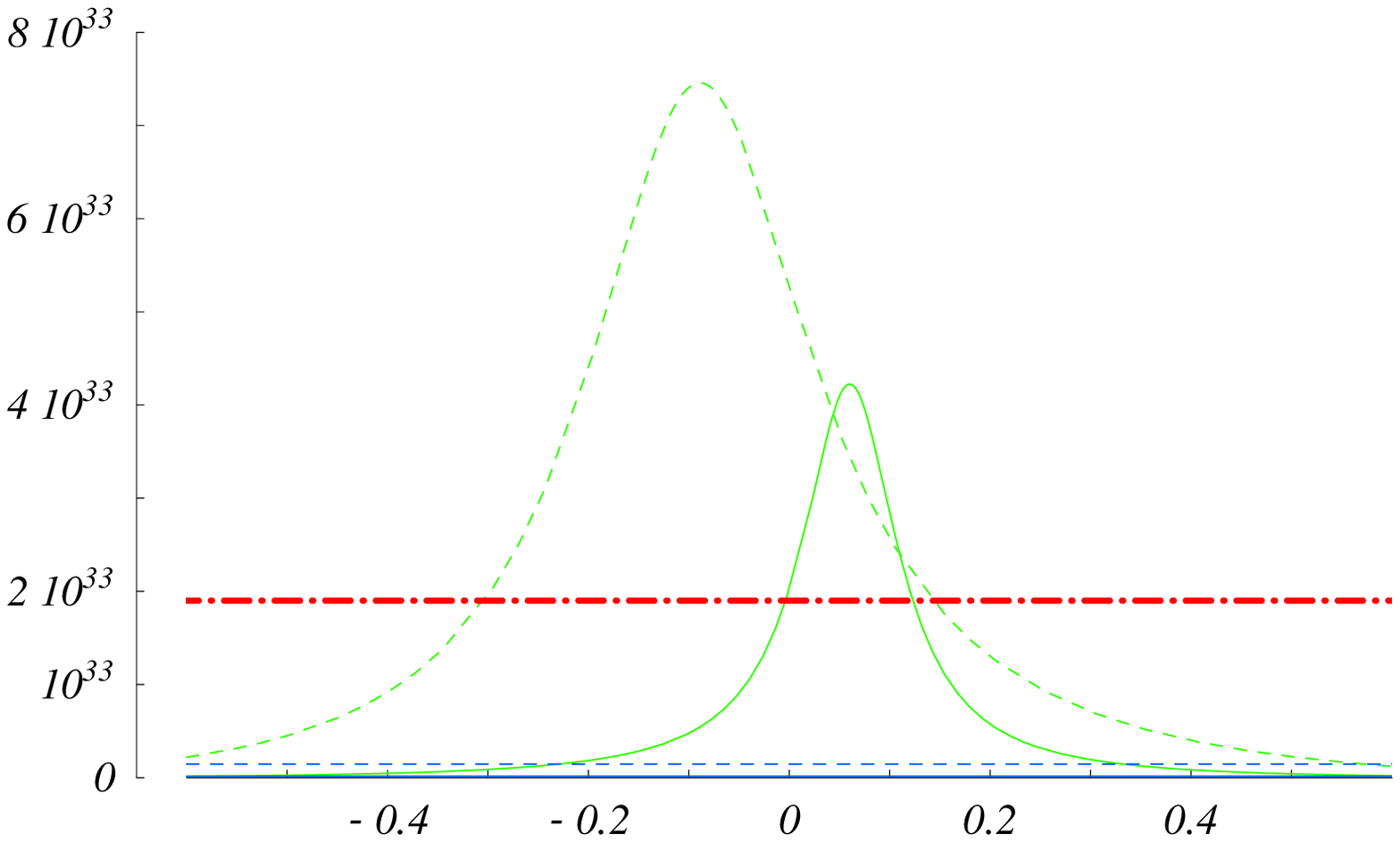,width=.7 \textwidth}
\put(-350,152){ \large $\tau_p$ [yrs] } 
\put(-60,-10){\large  $r$  } 
\put(-120,110){ $\longleftarrow$ (cpD)} 
\put(-113,103){ $\swarrow$} 
\put(-10,47){ $\uparrow$ exp. limit} 
\put(-0,15){ $ \leftarrow$ (deg) and (1t)  } }
\vskip 1.5 cm
\centerline{
\psfig{file=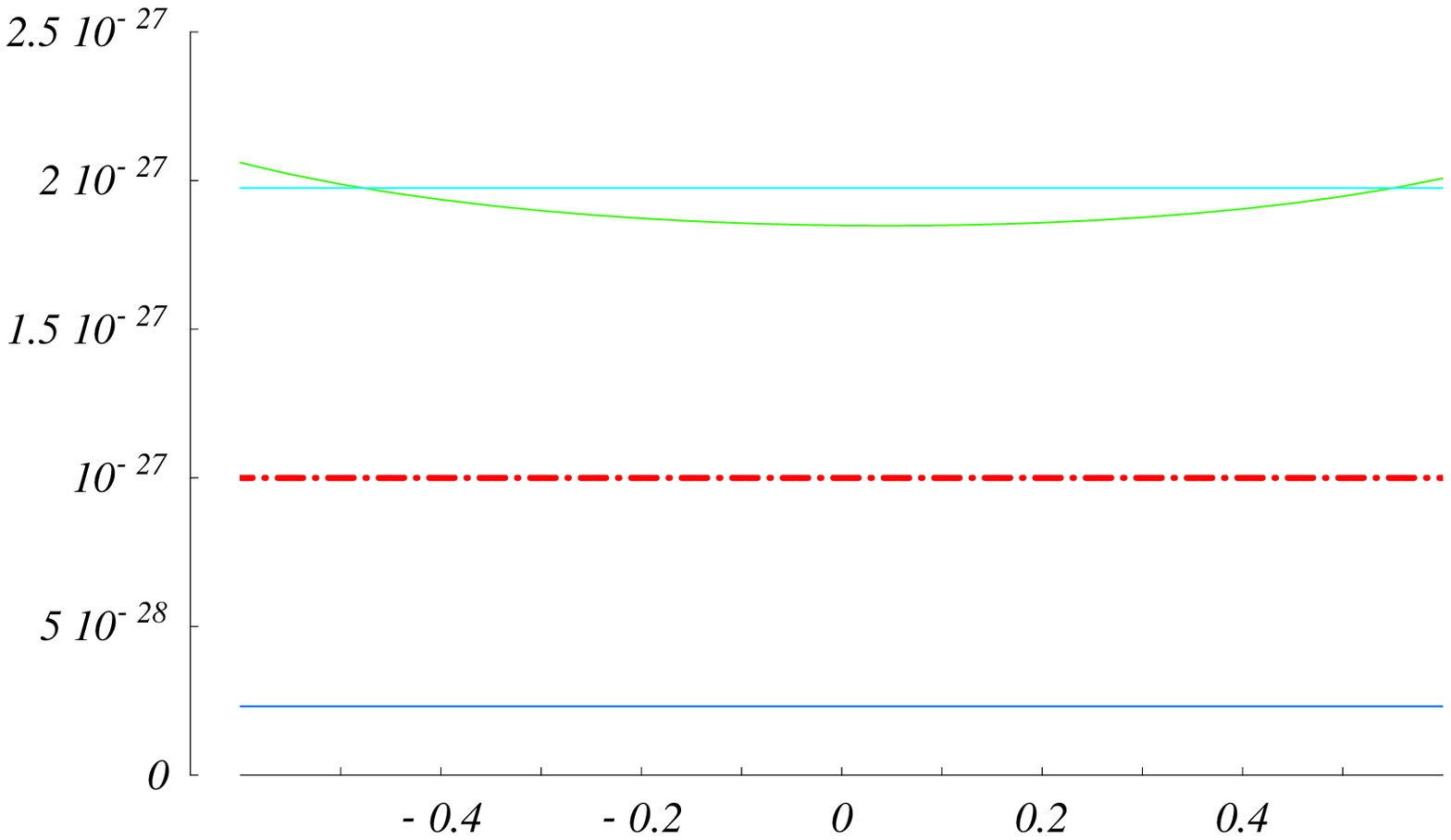,width=.7 \textwidth}
\put(-350,152){ \large $d_e$ [e cm] } 
\put(-60,-10){ \large $r$  }
\put(-105,110){ (cpD)} 
\put(-120,120){ $\nwarrow$} 
\put(-105,160){  (deg)}
\put(-120,150){ $\swarrow$} 
\put(-10,85){ $\downarrow$ exp. limit} 
\put(-0,30){ $ \leftarrow$ (1t)} }
\caption{
Predictions for $\tau_p$ and $d_e$ with three different choices for $M_T$: (1t), (deg) and (cpD),
which are defined in the text. We take 
$\bar M_T=10^{17}$ GeV, mSUGRA at point $P_2$ ($\ti M_1 =200$ GeV, $m_R=400$ GeV), 
$\tgb=3$. 
a) $\tau_p$: the dependence on the unknown phases in $\psi_u$ is also shown:  
for the solid line the phases maximize $\tau_p$, while for the dashed line the phases are set to zero. 
b) $d_e$: we take $\sin(2 \beta +\phi_1) = 1$, $M_3=\bar M_T$, $M_H=\bar M_T$ for (deg) and (cpD) 
while $M_H=M_{Pl}$ for (1t).}
\label{F_4t}
\vskip 1cm
\end{figure}

In this framework, assuming different patterns for $M_T$ and keeping fixed all the other parameters, 
let us now compare the corresponding predictions for $\tau_p$ and $d_e$.
In figs. \ref{F_4t} a) and b) we show the results of
the degenerate (deg) case, a class close to the pseudo-Dirac (cpD) case and the 
previously discussed one massive triplet (1t) case with $|{\cal C}_{31}| \sin\phi_{d_e} \approx 0.05
\sin(2 \beta + \phi_1)$, which represents the minimal $SU(5)$ limit:
\beq
M_T^{(deg)} = \mathbb{I} \bar M_T~,  \quad \quad
M_T^{(cpD)} = \left( \matrix{  r &  1 \cr 1 & r } \right) \bar M_T ~,\quad \quad 
M_T^{(1t)} = \left( \matrix{  \bar M_T &  0 \cr 0 &  M_{Pl} } \right) ~,
\eeq 
where $r$ is real and small free parameter ($r=0$ for pseudo-Dirac). 
As for the heavy doublet and heaviest right-handed neutrino masses
\footnote{The lighter right-handed Majorana masses need not to be specified, since they give negligible
contributions.},
for the degenerate and the pseudo-Dirac case we also set $M_H = M_3 = \bar M_T$, for the
one triplet case $M_H = M_{Pl}$ and $M_3 = \bar M_T$.
For definiteness, we choose $\tgb=3$, the sparticle masses at the point $P_2$,
$\bar M_T =10^{17}$ GeV.

In the minimal $SU(5)$ limit (1t), fig. \ref{F_4t}a) reproduces the well known result that the decay 
$p \ra K^+ \bar \nu$ comes out definitely too fast. 
The variation in the prediction due to the unknown phases contained in $\psi_u$ is also shown
(solid: phases maximising $\tau_p$; dashed: phases set to zero). 
Fig. \ref{F_4t}b), where $\sin(2 \beta +\phi_1) = 1$ has been taken, shows that the prediction for $d_e$ 
does not exceed $1/4$ of the experimental bound.

In the $SO(10)$ model with degenerate triplets (deg), 
$\tau_p$ is essentially unaffected 
as the new amplitudes are smaller than those already present in the minimal $SU(5)$.
As a consequence, fig. \ref{F_de-p}b) applies again.
On the contrary, since there are more states in the RGE, ${\cal I}_{11}$ gets enhanced with respect to (\ref{Irad}).
Its general expression when $\underline{2}$'s and $\underline{3}$'s are simultaneously diagonal is:
\beq 
{\cal I}_{11}  = {\cal I}m \left(    
y_u^\dagger  \left(
V_R^\dagger \ln\frac{\Lambda}{\hat M} V_R + 3 \ln \frac{\Lambda}{M_{T_2}}+\ln \frac{\Lambda}{M_H}
\right) y_u   \hat m_\ell 
y_u^T \left( \ln \frac{\Lambda}{M_{T_1}} + \frac{2}{3} \ln \frac{\Lambda}{M_H} \right) y_u^*  \right)_{11}.
\label{Irad2} 
\eeq
Hence, when all the heavy states are degenerate, $d_e$ is enhanced by a factor $25/3$ and,
as fig. \ref{F_4t}b) shows, it exceeds the experimental limit. 
Notice also that fig. \ref{Fde_comb} can be transposed to the (deg) case by multiplying $d_e^{(\tau_p)}$ 
by a factor $(25/3) \times (0.05/0.2) \approx 2$. So, for moderate $\tgb$, $d_e$ now becomes more restrictive 
on $M_T$ than $\tau_p$ for slepton masses up to $800$ GeV ($P_4$).

With the non-trivial structure $M_T^{(cpD)}$, triplets approach the pseudo-Dirac form 
as $r$ decreases and
their interference reduces the proton decay amplitude with the largest Yukawa couplings by a factor of $r,$ 
while the other amplitudes are disfavoured by their smaller couplings. 
This increases $\tau_p$ by up to two orders of magnitude if the phases $\psi_1$ and $\psi_2$ are
optimized, as shown in fig. \ref{F_4t}a). 
Instead, $d_e$ is only slightly affected by the change in the couplings and by $O(r^2/2)$ 
corrections due to the shift in the triplet eigenvalues 
(but not by the fact that in a pseudo-Dirac pair they have opposite CP phases). 
Notice that $M_T$ textures close to the pseudo-Dirac one 
have been widely used in the literature on realistic GUT models \cite{PD-lit} and
one should check whether such models also predict 
also $d_e$ in agreement with the present and planned experimental limits.

This numerical example shows that, 
in models where dimension 5 operators contributing to $\tau_p$ are suppressed 
by the choice of a rich structure for the triplet coupling and masses, 
the restrictions from the present limit on $d_e$ must be taken into account
and, \textit{a fortiori} the impact of future experimental improvements should be
evaluated. 
This is in spite of the relatively small interval of the RGE evolution, due to
the strong sensitivity of $d_e$ to the flavour and CP violations in the supersymmetric sector. 
Of course, the results crucially depend on the cutoff $\Lambda$ 
of the effective supersymmetric GUT theory, which in some special models 
could be below $M_{Pl}$ and suppress $d_e$ (see, e.g. the last work of ref. \cite{PD-lit}). 
Moreover, there are model dependent phases, like $\phi_1$ in our example, 
but generically there is no reason to believe that they should cancel 
with the other phases in $\phi_{d_e}$, so that  
the $d_e$ prediction essentially depends on $|{\cal C}_{31}|$.

\noindent \underline{ \large C) $d_e$  \textit{and LFV in neutrino mass models}}

To complete the analysis of this letter, let us first discuss for which values
of $|{\cal C}_{31}| \sin \phi_{d_e}$ the limits on $M_T$ from $d_e$ compare with those from $\tau_p$ 
and, secondly, let us look for the restrictions on $M_T$ in different classes of models in the literature.  

\begin{figure}[!b]
\vskip 1 cm
\centerline{\psfig{file=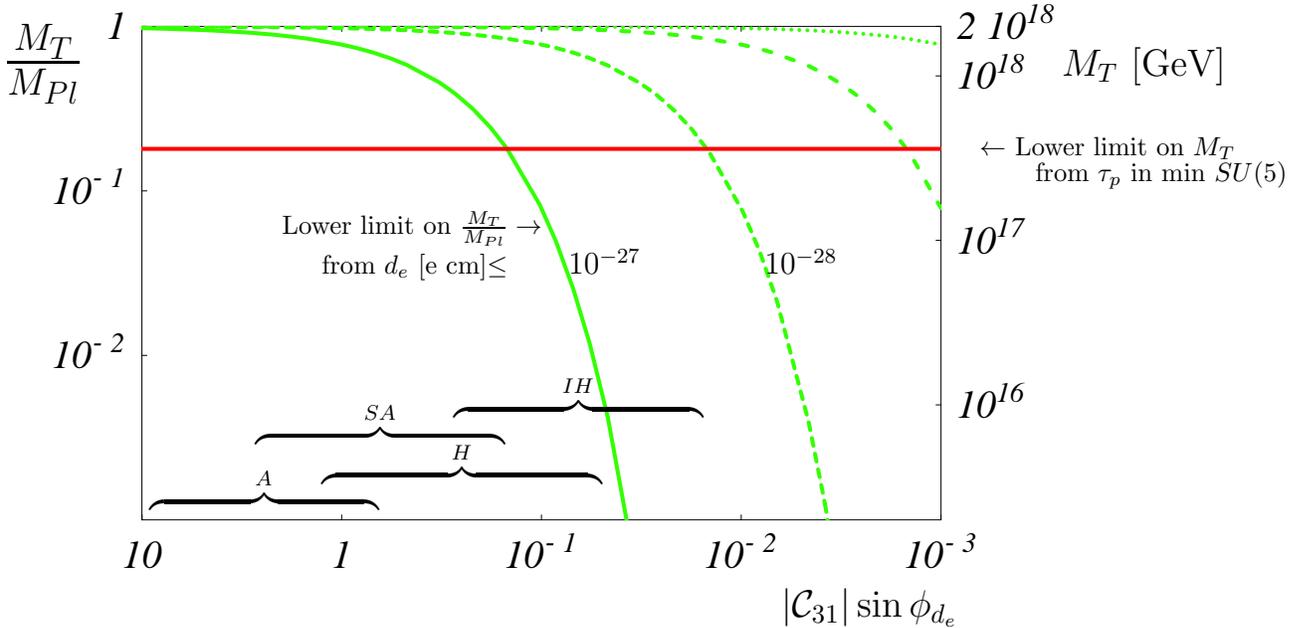,width=.9 \textwidth}
\put(0,200){\large $M_T$ [GeV]  }
\put(-35,170){ \footnotesize $\leftarrow$ Lower limit on $M_T$}
\put(-10,160){\footnotesize  from $\tau_p$ in min $SU(5)$}
\put(-400,200){\LARGE $\frac{M_T}{M_{Pl}}$}
\put(-295,140){\footnotesize Lower limit on  $\frac{M_T}{M_{Pl}}  \rightarrow$}
\put(-280,125){\footnotesize from $d_e $ [e~cm]$\le$ \normalsize$ ~~~~~10^{-27}$ ~~~~~~~~~~~$10^{-28}$ }
\put(-110,-5){ \large $|{\cal C}_{31}| \sin \phi_{d_e}$ }    
\put(-345,35){$\overbrace{~~~~~~~~~~~~~~~~~~~~~~}^{A}$ } 
\put(-305,60){$\overbrace{~~~~~~~~~~~~~~~~~~~~~~~~}^{SA}$ }
\put(-280,45){$\overbrace{~~~~~~~~~~~~~~~~~~~~~~~~~~~}^{H}$ } 
\put(-230,70){$\overbrace{~~~~~~~~~~~~~~~~~~~~~~~~}^{IH}$ } 
}
\caption{Lower limit on $M_T /M_{Pl}$ as a function of $|{\cal C}_{31}| \sin\phi_{d_e}$
for the minimal $SU(5)$ model with sparticles masses at $P_2$, $\tgb=3$. 
From left to right the curves refer respectively to $d_e < 10^{-27},  10^{-28},  10^{-29}$ e cm.
Also shown is the lower limit on $M_T$ from $\tau_p$, with $\psi_u$ phases chosen to maximize it.}
\label{Fde_comp}
\vskip 1cm
\end{figure}

This is shown in fig. \ref{Fde_comp}, where the lower limit on $M_T/\Lambda$ from $d_e$ 
is plotted for $\tgb=3$ and point $P_2$, for the present experimental sensitivity 
as well as for possible improvements by one and two orders of magnitude.
These results apply to the minimal $SU(5)$ case discussed in A) and  
can be quite easily adapted to more realistic cases. 
Indeed, $d_e$ increases with the addition of more triplets 
and this can be accounted for by rescaling the values on the horizontal axis of fig. \ref{Fde_comp}.
For instance, the (deg) and (cpD) models discribed in B)
correspond to the value $0.05 \times 25/3 \approx 0.4$ for $|{\cal C}_{31}| \sin\phi_{d_e}$. 
The lower limit on $M_T$ from $\tau_p$ in the case of minimal $SU(5)$ is also indicated.
With our choice of parameters, it turns out that $d_e$ presently supersedes $\tau_p$ for 
$|C_{31}| \sin\phi_{d_e}> 0.1$ and, as the experimental bound will be improved,
for proportionally smaller values. 

Let us now estimate the expectations for $|{\cal C}_{31}| \sin\phi_{d_e}$ in different neutrino mass models 
in the literature. For the present discussion, they can be divided into two categories:
a) models where $|{\cal C}_{31}|$ and $|{\cal C}_{21}|$ are naturally of the same order of magnitude; 
b) models where  $|{\cal C}_{31}| < |{\cal C}_{21}|$. 
Indeed, the limit on $\mu \ra e \gamma$ implies an upper limit on $|{\cal C}_{21}|$ 
which, for point $P_2$, corresponds to $|{\cal C}_{21}| \le 0.1 \times 3/\tgb$ \cite{imsusy02}.  
In the following we consider some typical examples of textures in the case that $y_{\nu_3} = y_{t}$. 

Examples of models in category a) are the $U(1)$-flavour symmetry models 
compatible with $SU(5)$ studied in ref. \cite{afm3},
which we refer to for the details and for proper references in the literature.  
Such textures were classified according to their amount of structure as:  
anarchical (A), semi-anarchical (SA), hierarchical (H) and inversed-hierarchical (IH).
The corresponding expectations for $|{\cal C}_{31}| \sin \phi_{d_e}$ are displayed in fig. \ref{Fde_comp}.  
These are only generic order of magnitude predictions because of the nature of the models 
and the uncertainty in $\sin \phi_{d_e}$. 
As apparent from fig. \ref{Fde_comp}, 
only (H) and (IH) models do not conflict with the experimental bound on $\mu \ra e \gamma$ 
but they require quite a very high $M_T$, above $10^{17(18)}$ GeV with $d_e < 10^{-27(-28)}$ e cm, 
which is comparable to (stronger than) the lower bound from $\tau_p$ in minimal $SU(5)$. 
However, as already stressed, while $\tau_p$ is sensitive to the couplings of the lighter generations 
and could significantly change if the minimal relations in eq. (\ref{minsu5}) are relaxed, 
on the contrary $d_e$ depends on the third generation Yukawa couplings and should be slighly affected. 

Category b) includes models with small mixing angles for $y_\nu$. 
Consider first an $SU(5)$ model where $V_L \approx V_{CKM}$ 
(there is no conflict with $\mu \ra e \gamma$ since, at $P_2$, 
$|{\cal C}_{21}| \approx 2~10^{-3} \times 3/\tgb$).  
As already discussed in A), in such model $|{\cal C}_{31}| \sin \phi_{d_e} \sim 0.05 \sin(50^{\circ} + \phi_1)$, 
which is naturally $O(10^{-2})$. 
At present this is compatible with $M_T \sim M_{GUT}$, but a limit on $d_e$ 
at the level of $10^{-28}$ e cm would require $M_T \gg M_{GUT}$. 
Particularly interesting are the $SO(10)$-inspired models, where $y_{\nu_3} = y_{t}$ comes out as a prediction.  
As already mentioned, the (deg) and (cpD) cases of the model with two $\underline{10}$'s discussed in B) 
correspond to a value $0.4$ in abscissa. Since the $d_e$ bound is  
satisfied only with an unnaturally large value for $M_T$, these two specific models are excluded. 
Notice also that models with non-abelian $U(2)$ or $SU(3)$ flavour symmetries usually fall into category b). 
In explicit models \cite{rossvelasco}, ${V_L}_{\tau 1}$ could be even smaller than $V_{td}$ and 
to estimate $d_e$ one should inspect the number of states involved in the RGE. 
We point out that the present and planned experimental limits on $d_e$ 
are privileged tools to check and eventually disprove such non-abelian flavour symmetries models.
 
From the above analysis (see also the plots in ref. \cite{isa-edm}) it turns out that, 
if the triplet masses are reasonably assumed to be at the gauge coupling unification scale, 
$M_T\sim M_{GUT}$, the present $d_e$ experiments are already at the edge of testing
the range of $|{\cal C}_{31}| \sin \phi_{d_e}$ values that are predicted in grand-unified neutrino mass models. 
Hence, future searches for LFV decays and $d_e$ 
will provide many different constraint on the neutrino mass sector of these models. 

%%%%%%%%%%%%%%%  Conclusions  %%%%%%%%%%%%%%%%%%%%%%%%%%%%%%%

\noindent \underline{ \large \textit{Concluding remarks}}

In supersymmetric grand-unified theories, important contributions to $d_e$ are associated
to the simultaneous violations of lepton flavour and CP 
in the Yukawa couplings of the colour triplet partners of the Higgs doublets 
and in those of the right-handed neutrinos.  
In this paper we have carried out
a comparison between the estimate of these effects \cite{isa-edm} and
the predictions for the proton lifetime,
proving that both experiments are quite competitive in putting limits on the colour triplet masses,
hence on the pattern of supersymmetric GUTs.
Actually, $d_e$ turns out to be more effective in two respects: 
it increases with the number of triplets and is quite insensitive to the triplet mass matrix structure 
that is on the contrary crucial to suppress proton decay.  
Therefore, $d_e$ bounds should be carefully checked in potentially 
realistic supersymmetric GUT models.  
Moreover, $d_e$ depends on a piece of the neutrino mass puzzle of difficult experimental access 
- it is related to the decay $\tau\rightarrow e \gamma$ - 
which, as shown here, should be effectively constrained by the future searches for $d_e$. 
Neutrino mass models are thus directly concerned by this constraint.

%%%%%%%%%%%%%%%  Acknowledgements  %%%%%%%%%%%%%%%%%%%%%%%%%%%

{\it Acknowledgements:} The authors are grateful to the Dept. of Physics of "La Sapienza" University in Rome 
for the warm hospitality during the completion of this work.   
This work has been supported in part by the RTN European Program HPRN-CT-2000-00148.

%%%%%%%%%%%%%%   THE   END  % %%%%%%%%%%%%%%%%%%%%%%%%%%%%%%%

%%%%%%%%%%%%%%  Bibliography   %%%%%%%%%%%%%%%%%%%%%%%%%%%%%%%

\end{document}